\documentstyle[12pt]{article}
\input{psfig.tex}
\newcommand{\etal}{{\em et\ al.}\ }

\newcommand{\kmsec}{km s$^{-1}$\ }
\newcommand{\kms}{km s$^{-1}$\ }

\newcommand{\chisq}{${\chi}^{2}$}

\newcommand{\Ho}{H$_{0}$}
\newcommand{\qo}{q$_{0}$}
\newcommand{\kmsMpc}{km s$^{-1}$ Mpc$^{-1}$\ }

\def\lesssim{\mathrel{\hbox{\rlap{\hbox{\lower4pt\hbox{$\sim$}}}\hbox{$<$}}}}
\def\gtrsim{\mathrel{\hbox{\rlap{\hbox{\lower4pt\hbox{$\sim$}}}\hbox{$>$}}}}

\begin{document}

\vspace{0.1in}

\centerline{                       THE HUBBLE DIAGRAM OF}
\centerline{                   THE CAL\'{A}N/TOLOLO TYPE Ia SUPERNOVAE}
\centerline{                   AND THE VALUE OF \Ho}

\vspace{0.4in}
\centerline{                               Mario Hamuy$^{1,2}$}
\centerline{                              M. M. Phillips$^1$}
\centerline{                           Nicholas B. Suntzeff$^1$}
\centerline{                            Robert A. Schommer$^1$}
\centerline{                                 Jos\'{e} Maza$^3$}
\centerline{                                 R. Avil\'{e}s$^1$}
\vspace{0.4in}
\noindent $^1$ National Optical Astronomy Observatories$^*$,
Cerro Tololo Inter-American Observatory, Casilla 603, La Serena, Chile\\

\noindent $^2$ University of Arizona, Steward Observatory, Tucson,
Arizona 85721\\

\noindent $^3$ Departamento de Astronom\'{i}a, Universidad de Chile, Casilla 36-D,
Santiago, Chile\\

\noindent electronic mail: mhamuy@as.arizona.edu, mphillips@noao.edu,\\
\noindent nsuntzeff@noao.edu, rschommer@noao.edu, jmaza@das.uchile.cl\\
\vspace{0.4in}

\noindent Running Page Head : HUBBLE DIAGRAM OF SNe Ia\\

\noindent Address for proofs:  M. M. Phillips, CTIO, Casilla 603, La Serena, Chile\\

\vspace{0.1in}

\footnoterule

\vspace{0.1in}

\noindent $^*$Cerro Tololo Inter-American Observatory, National
Optical Astronomy Observatories, operated by the Association of Universities
for Research in Astronomy, Inc., (AURA), under cooperative agreement with
the National Science Foundation.\\

\eject
\centerline{                        ABSTRACT}

     The Cal\'{a}n/Tololo supernova survey has discovered $\sim$30 Type Ia
supernovae at redshifts out to z$\sim$0.1. Using BV(I)$_{KC}$ data for these objects
and nearby SNe Ia, we have shown that there exists a significant dispersion
in the intrinsic luminosities of these objects. We have devised a robust \chisq
minimization technique simultaneously fitting the BVI light curves to parametrize
the SN event as a function of (t$_{B}$, m$_{i}$, $\Delta$m$_{15}$(B)) where
t$_{B}$ is the time of B maximum, m$_{i}$ is the peak BVI magnitude corrected
for luminosity variations, and $\Delta$m$_{15}$(B) is a single parameter describing
the {\it whole} light curve morphology. When properly corrected for $\Delta$m$_{15}$(B),
SNe Ia prove to be high precision distance indicators, yielding
relative distances with errors $\sim$7-10\%. 
The corrected peak magnitudes are used to construct 
BVI Hubble diagrams, and with Cepheid distances recently measured with the Hubble
Space Telescope to four nearby SNe Ia (1937C, 1972E, 1981B, and 1990N) we derive 
a value of the Hubble constant of \Ho=63.1$\pm$3.4 (internal) $\pm$2.9 (external) \kmsMpc.
This value is $\sim$10-15\% larger than the value obtained by assuming that SNe Ia
are perfect standard candles. 
As we have shown in Paper V, there is
now strong evidence that galaxies with younger stellar population (spirals and
irregulars) appear to host the slowest-declining, and therefore most luminous SNe Ia.
Hence, the use of Pop I objects such as Cepheids to calibrate the zero point of
the SNe Ia Hubble diagram can easily bias the results toward luminous SNe Ia,
{\it unless the absolute magnitude-decline relation is taken into account.}

Using the inertial reference frame of the distant SNe Ia,
and the "corrected" apparent magnitudes of the best observed SNe Ia in the Virgo
and Fornax clusters,
we have evaluated the recession velocities of these two clusters based on the Cal\'{a}n/Tololo
Hubble diagrams. We find a cosmic recession velocity for Virgo of 1,223$\pm$115 \kms and
for Fornax of 1,342$\pm$70 \kms, and a relative distance modulus of  
$\Delta\mu$(Fornax-Virgo) = 0.19$^m$$\pm$0.23$^m$.

\eject

\section{  Introduction}
     In mid-1990, a group of astronomers of the Cerro Tololo Inter-American
Observatory (CTIO) and the University of Chile at Cerro Cal\'{a}n initiated
a photographic search for supernovae with the aim of producing a moderately
distant (0.01 $\lesssim$ z $\lesssim$ 0.1) sample suitable for cosmological
studies. The technique employed in our survey was described in detail by
Hamuy \etal 1993a (Paper I). In the course of 1990-93 this project led to
the discovery of 50 SNe ($\sim$1/3 of all of the SNe discovered
throughout the world in the same period). Followup spectroscopic observations
revealed that 32 of these objects were members of the Type Ia class, the
majority of
which were found near maximum light.

Thanks to the generous collaboration
of many visiting astronomers and CTIO staff members, we were able to gather
BV(RI)$_{KC}$ CCD photometry for 27 of these Cal\'{a}n/Tololo SNe Ia. 
Preliminary light curves and spectroscopy were published in Paper I, 
Paper II (Maza \etal 1994), and Paper III 
(Hamuy \etal 1994) for six events. Based on photometry 
for 13 of the Cal\'{a}n/Tololo SNe Ia, in Paper IV (Hamuy \etal 1995) we 
discussed the Hubble diagrams in B and V.  The latter study confirmed,
in general terms, the finding by Phillips (1993) that the absolute
B and V magnitudes of SNe Ia are correlated with the initial decline rate
of the B light curve.  In addition, we found evidence that galaxies having a
younger stellar population appear to host the most luminous SNe~Ia, and
pointed out that this effect could introduce significant bias into 
determinations of both $H_o$ and $q_o$.  A significant decrease in the
scatter of the B and V Hubble diagrams was obtained when the data were
corrected for the peak luminosity-decline rate relation, and a Hubble
constant in the range $H_o$ = 62--67 km s$^{-1}$ Mpc$^{-1}$ was found
using the Cepheid distance to the host galaxy of SN~1972E to derive the
zero point of the Hubble relation.

We have now produced definitive light curves for the complete sample of 
27 Cal\'{a}n/Tololo SNe Ia, and also for
two additional SNe Ia found in the course of other survey programs which we
included in our followup program.  
In an accompanying paper (Hamuy \etal 1996a; hereafter refered to as Paper V), 
we have reexamined the absolute luminosities of
this full set of 29 SNe~Ia.  These data confirm the major conclusions of 
Paper IV concerning 1) the reality of the absolute magnitude-decline rate
relation, and 2) the link between SNe~Ia luminosities and the morphology
of the host galaxies.
The purpose of the current paper is to present and discuss the final versions of the
Hubble diagram in B and V for the full set of Cal\'{a}n/Tololo SNe Ia, as well as the first
Hubble diagram in the I band.  After a brief description of our sample (Sec. 2),
we present in Sec. 3 the resulting BVI Hubble diagrams. In Sec. 4 we discuss
our results, and in Sec. 5 we summarize our conclusions.

\section{  The Cal\'{a}n/Tololo Database}

     Of the 32 SNe Ia discovered during the course of the Cal\'{a}n/Tololo
survey, adequately sampled
BV(I)$_{KC}$ light curves were obtained for a total of 27 events.  To this sample we have added two
SNe Ia -- 1990O and 1992al -- which, although not found by us, were included in our
program of followup photometry.  Hence, the total sample of distant SNe Ia considered in this
paper is 29.

In an accompanying paper (Hamuy \etal 1996b;
hereafter refered to as Paper VII), we give
the individual BV(RI)$_{KC}$ light curves for this set of 29 SNe Ia
along with details of the photometric reductions.  Estimates of the
maximum light magnitudes in the B, V, and I bands and the initial
decline rate parameter $\Delta$m$_{15}$(B) (Phillips 1993) are also
provided.  As discussed in Paper VII, 
maximum-light magnitudes were measured directly from the observations whenever
possible. We were able to perform this measurement for 11 SNe for which the time of the first photometric
observation was no later than day +1 (counted since the peak of the B light curve), by
fitting a low-order (3-4) polynomial to the data around maximum light.
For the remaining 18 objects whose followup photometric observations 
did not begin until after maximum light, the peak magnitudes
were estimated using a robust \chisq-minimizing fitting procedure similar to that
described in Paper IV. Note, however, that in order to include the I-band data, an updated
and expanded set of BV(I)$_{KC}$ light curve templates were employed.  This new set of templates
are the subject of Paper VIII (Hamuy \etal 1996c) in this series.
These templates were fit to the observed photometry
of each SN solving simultaneously  for the time of B maximum and the peak magnitudes
B$_{MAX}$, V$_{MAX}$, and I$_{MAX}$. (See Paper VII for further details of this fitting 
procedure.)

In Table 1 we list the relevant information for this study for the 29 SNe in the following
format:\\

\noindent Column (1): SN name\\
\noindent Column (2): decimal logarithm of the heliocentric radial velocity (cz) of
the parent galaxy corrected to the cosmic microwave background (CMB)
frame. The heliocentric redshifts were all
measured from our own spectroscopic observation of the nuclei of the parent
galaxies and the transformation to the CMB frame was performed by adding the
vectors (-30,297,-27) and (7,-542,302) {\kmsec} (in Galactic Cartesian Coordinates).
The former is the motion of the sun with respect to the Local Group (Lynden-Bell
\& Lahav 1988) while the latter is the motion of the Local Group with respect
to the CMB as measured with COBE (Smoot \etal 1992). We assigned a possible
peculiar velocity component of $\pm$600 {\kmsec} to each galaxy which is given
in brackets in units of mdex.\\
\noindent Column (3), (4) and (5): the apparent B, V, and I peak magnitudes of
the SN as determined from the light curves, corrected for foreground extinction
in the direction of the host galaxy (Burstein \& Heiles 1982), and for the K terms
calculated by Hamuy \etal (1993b).\footnote[1]
{ Since the K terms for SNe Ia in the I band are not available in the literature
we calculated these corrections based on our own spectroscopic observations
of SNe Ia gathered at CTIO, following the same precepts described by Hamuy \etal (1993b).}
{ \it Note, however, that no correction has been applied for
possible obscuration in the host galaxy} (see Sec. 3).\\
\noindent Column (6): the ``color'' of the SN, B$_{MAX}-$V$_{MAX}$.\\
\noindent Column (7), (8) and (9): the absolute B, V, and I peak magnitudes of
the SN calculated from the apparent magnitude, the redshift of the host galaxy
(in the CMB frame), and an assumed
Hubble constant of H$_\circ$= 65 \kms Mpc$^{-1}$. As discussed in Sec. 4
this value corresponds closely to the Hubble diagram of SNe Ia with its
zero point duly calibrated with Cepheid distances to nearby SNe Ia.\\
\noindent Column (10): the decline rate parameter $\Delta$m$_{15}$(B), defined by
Phillips (1993) as the amount in magnitudes that the B light curve decays in
the first 15 days after maximum light. In most cases (24 SNe) this parameter was
estimated through the template fitting procedure previously described in Papers III and IV.\\
\noindent Column (11): an estimate (in Angstroms) of the equivalent width of Na I D
interstellar absorption due to the host galaxy from the spectra of the SNe.
In most cases we were only able to place an upper limit to the equivalent width based
on the signal-to-noise ratio of the spectra.\\

\section{ The Hubble Diagrams in B, V, and I}

     Figure 1 shows the resulting Hubble Diagrams in B, V, and I for the  Cal\'{a}n/Tololo SNe Ia
data given in Table 1. The ridge lines, which correspond to (weighted) least-squares fits
of the data to the theoretical lines of constant luminosity, yield the following
zero points and dispersions:

\vspace{0.2in}
\centerline{         B$_{MAX}$ = 5 log cz - 3.151 ($\pm$0.029)     $\sigma$ = 0.38 n = 29,   (1)}
\vspace{0.2in}
\centerline{         V$_{MAX}$ = 5 log cz - 3.175 ($\pm$0.025)     $\sigma$ = 0.26 n = 29,   (2)}
\vspace{0.2in}
\centerline{         I$_{MAX}$ = 5 log cz - 2.948 ($\pm$0.029)     $\sigma$ = 0.19 n = 25.   (3)}

\noindent Note that the quoted dispersion is the simple {\it rms} deviation of the points about
the fit.

The most obvious property of these diagrams is that the dispersion is greatest for the
nearest SNe.
As pointed out in Paper IV, this property
would be expected if the scatter were due primarily to the peculiar motions of the host galaxies.
This would require that the galaxies possess peculiar motions with respect
to the pure Hubble flow considerably in excess of the value of 600 \kmsec assumed in
the horizontal error bars plotted in Figure 1. If we interpret the entire
scatter in the V data in this figure as due to peculiar velocities, the {\it rms} velocity
scatter would be $ <(v_{CMB} - v_{Hubble flow})^2>^{1/2} $ = 1420 \kmsec. This
value is dominated by large residuals at higher redshift; however, if we
restrict the sample to the 11 galaxies within 10,000 \kmsec, the {\it rms} residual
velocity is still 1050 \kmsec. These
values are considerably in excess of measurements of the  cosmic velocity
dispersion. For example, Marzke \etal (1995) measure  pairwise velocity differences from
the CfA2 and Southern Sky Redshift Survey  ranging between 299 $\pm$ 99 \kmsec and 540 $\pm$
180 \kmsec; the larger values are derived when dense virialized systems are included in
the analysis. Since the SN found in the Cal\'an/Tololo sample avoid rich Abell
clusters and the scatter in the Hubble diagram derived above are still 
3$\sigma$ in excess of these
larger values for pairwise velocities, it is extremely unlikely that a
dominant portion of the observed scatter arises from  peculiar motions.

Dust in the parent galaxies probably also contributes to the observed dispersion
in the Hubble diagrams. Indeed, Table 1 shows that Na I D interstellar lines were
detected in the spectra of a few of the SNe. In order to examine this alternative,
we plot in Figure 2 the absolute magnitudes of the Cal\'{a}n/Tololo SNe as a
function of the B$_{\rm{MAX}}-$V$_{\rm{MAX}}$ color. Plotted for comparison in the
same figure are the Galactic reddening vectors (with slopes of 4.1 in B, 3.1 in V,
and 1.85 in I) corresponding to a color excess of E(B-V)=0.2.  We see that 26 of
the 29 SNe lie in a well-defined clump centered on
B$_{\rm{MAX}}$-V$_{\rm{MAX}} \sim 0.0$.
Of the 3 outliers, SN~1992K is a SN~1991bg-like event whose very red color
and low luminosity is almost certainly intrinsic (see Paper III).
The spectrum of SN~1993H showed weak Na~I~D absorption (see Table 1), 
and so its red color 
may be evidence of significant dust extinction.  However, the maximum-light
spectrum of this object also showed Ti II absorption, similar in strength to that 
observed in the fast-declining SN~1986G (Phillips \etal 1987), which may
indicate that the red color of this SN was at least partly intrinsic
(see Nugent \etal 1995).  In the case of SN~1990Y, the one spectrum available
to us is of such low signal-to-noise that we can
only place an uninteresting upper limit of 4 \AA  on the equivalent width of any
possible absorption due to the host galaxy.  Hence, we cannot say whether the
red color of this SN was intrinsic or due to dust extinction.
The mean color of the remaining 26 SNe is B$_{\rm{MAX}}$-V$_{\rm{MAX}}$ = 0.03, with an
{\it rms} dispersion of only 0.06 mag.  If this dispersion were
entirely due to dust extinction in the host galaxies, it could account for
as much as 0.25$^{m}$ of the dispersion in the B Hubble diagram, 0.19$^{m}$ in V, 
and 0.11$^{m}$ in I.  However, it is likely that at least part of the observed 
dispersion in color is intrinsic (e.g., SNe 1992bc versus 1992bo; Paper II) which,
coincidentally, appears to run in the same direction as the reddening vectors.
Although sorting out these two different possibilities --dust reddening
versus an intrinsic color spread-- is made difficult by the similar effects
they display in a diagram like Fig. 2,
we conclude that dust extinction in the parent galaxies is
generally small, and cannot explain all of the observed dispersion in
the Hubble diagrams.

As shown in Paper V, the major source of the dispersion in the Hubble diagrams
in Figure 1 is an intrinsic spread in the peak luminosities of SNe~Ia.
Fortunately, as shown by Phillips (1993) and confirmed in Paper V (and also in
a preliminary way in Paper IV), the absolute
magnitudes at maximum light are strongly correlated with the initial decline rate of
the B light curve. The sense of this correlation is that the most luminous SNe~Ia
display the slowest decline rates.  In Figure 1 of Paper V, the absolute magnitudes 
of the Cal\'{a}n/Tololo SNe are plotted as a function of the decline rate parameter
$\Delta$m$_{15}$(B).  Excluding the 3 events (SNe 1990Y, 1992K, and 1993H) in the sample
with B$_{\rm{MAX}}$-V$_{\rm{MAX}} >$ 0.2, linear least-square fits to the data for the 26
remaining SNe yield slopes 
of 0.784$\pm$0.182 in B, 0.707$\pm$0.150
in V and 0.575$\pm$0.178 in I.  We shall use these relations in Sec. 4 to correct
the Hubble diagrams for this effect.

A magnitude limited SNe search such as the Cal\'{a}n/Tololo survey is
susceptible to variety of possible selection effects. Lower luminosity 
events will be increasingly difficult to find at larger
redshifts, leading to classical Malmquist bias.  In order to study this effect,
we plot in Figure 3 the absolute BVI magnitudes of the Cal\'{a}n/Tololo SNe
as a function of redshift.  The most distant SNe (log cz
$>$ 4) are not significantly brighter than the nearby sample, although it is
clear that the faintest of the nearby SNe are not found in the distant sample.
Indeed, the mean values for the nearest 11 SNe are slightly brighter than the
distant sample of 18 (M$^B_{\rm{MAX}}$= -19.08 $\pm$ 0.17  for the nearby sample vs.
$-19.04 \pm 0.06$ for the distant SNe.)  The major effect seen in Figure 3 
is the larger scatter at lower redshifts.  Malmquist bias may, however, affect
the {\em slope} of the absolute magnitude-decline rate relation.  In Paper V we
found that the slope obtained for a nearby sample of SNe~Ia with Cepheid, Surface
Brightness Fluctuations, or Planetary Nebula Luminosity Function distances
appeared to be steeper than the slope derived for the Cal\'{a}n/Tololo sample.
We note that the four intrinsically-faintest SNe in the nearby sample would not
have been discovered over most of the volume surveyed by the Cal\'{a}n/Tololo
search, and thus the slopes obtained for the Cal\'{a}n/Tololo sample could
be biased to flatter values.

A puzzling aspect of the sample which contributes to the scatter in this figure is 
that SNe hosted by spiral galaxies, which are the most luminous events of our sample
(see Paper V), are preferentially found in the nearer sample. This apparent selection
effect can be appreciated more clearly in the upper half of Figure 4 which shows the morphological
types of the host galaxies of the Cal\'{a}n/Tololo SNe Ia (separated into basic
categories of spirals and nonspirals) plotted as a function of redshift. Clearly
the ratio of spirals/nonspirals is significantly higher in the nearby sample
(log cz $<$ 4) than in the more distant group (log cz $>$ 4). In the lower half of Figure 4 we show
another representation of this effect in a plot of $\Delta$m$_{15}$(B) vs redshift.
In the nearby sample we find  three of the four largest values of
$\Delta$m$_{15}$(B), as might be expected from a Malmquist effect, 
since these are the fainter SNe. But
we also find the four smallest values of $\Delta$m$_{15}$(B) 
in this nearby
sample, representing the brightest SNe. Thus the overall effect is to
produce a large scatter in the Hubble diagram for nearby SNe. 

As briefly discussed in Paper IV, there are several possible causes for this effect. It could be
that the SNe in spiral galaxies occur preferentially near the central parts of the host
galaxies or near spiral arms, posing increasing difficulties to their
discovery at larger redshifts in a photographic search. There may also
have been a  bias in the survey to avoid nearby  clusters (e.g., Virgo or Fornax)
thus preferentially sampling the spiral rich field nearby; at larger distances
rich clusters might not have been avoided, thus giving the observed
morphological signal. While these are plausible explanations, the definitive
cause of this rather unusual sample effect is still unclear. Whatever the
exact cause, it appears that the derivation of the Hubble constant will
not be greatly affected by this somewhat bizarre sample effect because we
correct for the dependence of absolute magnitude on $\Delta$m$_{15}$(B). We caution,
however, that attempts to derive the SNe Ia luminosity function from
these data clearly {\it would} be affected, and the selection effects
warrent more careful consideration.
 
\section{Discussion}

\subsection{ The corrected Hubble diagrams}

The existence
of the absolute magnitude-decline rate relation allows us to correct the
apparent magnitudes of the Cal\'{a}n/Tololo SNe and reanalyze the corresponding
Hubble diagrams. In keeping with Paper V, we shall limit the sample to the 26
events with B$_{\rm{MAX}}-$V$_{\rm{MAX}}$ $\leq$ 0.20.\footnote[2]{A color cutoff
(at B$_{\rm{MAX}}-$V$_{\rm{MAX}} <$ 0.25) as an objective criteria to exclude
heavily-reddened or intrinsically low-luminosity events has been proposed by 
Vaughan \etal (1995).}
As mentioned in Sec. 3 there is little room for significant extinction in the
parent galaxies of this subsample of 26 events.
Linear regression fits to the data for this subsample (ignoring for the moment the
absolute magnitude-decline rate dependence) give the following
zero points and dispersions:

\vspace{0.2in}
\centerline{         B$_{MAX}$ = 5 log cz - 3.177 ($\pm$0.029)     $\sigma$ = 0.24 n = 26,   (4)}
\vspace{0.2in}
\centerline{         V$_{MAX}$ = 5 log cz - 3.188 ($\pm$0.026)     $\sigma$ = 0.22 n = 26,   (5)}
\vspace{0.2in}
\centerline{         I$_{MAX}$ = 5 log cz - 2.958 ($\pm$0.030)     $\sigma$ = 0.19 n = 22.   (6)}

\noindent The upper half of Figure 5 shows the V Hubble
diagram for this subsample of 26 SNe.  Excluding the 3 reddest
events has helped to decrease the dispersion [cf. Equations. (1)--(3)], especially
in the B band (0.24$^{m}$ versus 0.38$^{m}$ for the full sample).  The effect in I,
on the other hand, is negligible.

A more dramatic decrease in the dispersion of the Hubble diagrams is obtained by
correcting for the peak luminosity-decline rate relation.
Adopting the slopes derived in Paper V from the same sample of 26 Cal\'{a}n/Tololo 
SNe~Ia, we obtain:\\ 

\vspace{0.2in}
\centerline{B$_{MAX}$ - 0.784($\pm$0.182) [$\Delta$m$_{15}$(B) - 1.1] = 5 log cz - 3.318 ($\pm$0.035)     $\sigma$ = 0.17 n = 26,   (7)}
\vspace{0.2in}
\centerline{V$_{MAX}$ - 0.707($\pm$0.150) [$\Delta$m$_{15}$(B) - 1.1] = 5 log cz - 3.329 ($\pm$0.031)     $\sigma$ = 0.14 n = 26,   (8)}
\vspace{0.2in}
\centerline{I$_{MAX}$ - 0.575($\pm$0.178) [$\Delta$m$_{15}$(B) - 1.1] = 5 log cz - 3.057 ($\pm$0.035)     $\sigma$ = 0.13 n = 22.   (9)}

\noindent The second term on the left of these equations corrects the observed magnitudes
of each SN to the equivalent magnitudes of an event with $\Delta$m$_{15}$(B) = 1.1 mag.
The lower half of Figure 5 shows the corrected V Hubble diagram.
Application of the peak magnitude-decline rate relation has
reduced the scatter in this diagram to a level of 0.14$^m$ which, as
the value of the reduced $\chi$$^{2}$ indicates, is now entirely consistent with the assumed 
errors.  The reduction of $\sigma$ by approximately a factor of 1.5 allows SNe~Ia to be
used as excellent distance indicators (with precisions in relative distances $\sim$7-10\%).
If we, again, interpret this scatter as being entirely due to peculiar velocities, the
11 nearest objects imply an rms velocity of 550 \kms.  A more detailed analysis of the
velocity field based on these data will be given in a separate paper (Suntzeff \etal 1996a).\\

We can test the sensitivity of these results to dust extinction in the host galaxies
by restricting the sample even further to those objects for which we have been able to place 
an upper limit to the equivalent width of the Na I D interstellar lines $\leq$0.5~\AA.
For gas typical of that found in the disk of our own Galaxy, such a limit corresponds
to color excesses E(B-V) $\lesssim$ 0.1.
For the 15 SNe meeting this condition (90O, 90af, 91S, 91ag, 92J, 92ae, 92al, 92bc, 92bg,
92bh, 92bk, 92bl, 92bo, 93B, 93O), the zero points of the corrected Hubble diagrams prove
to be -3.318$\pm$0.046 in B, -3.324$\pm$0.041 in V, and -3.039$\pm$0.046 in I, which are
insignificantly different than the zero points obtained from the 26 SNe.

\subsection{ The Hubble constant}

Recently, Sandage and collaborators have measured Cepheid distances to the host galaxies
of six nearby SNe Ia (SN 1937C in IC 4182, SNe 1895B and 1972E in NGC 5253, SN 1981B in NGC 4536,
SN 1960F in NGC 4496, and SN 1990N in NGC 4639) with the aim to calibrate the Hubble diagram
of the distant SNe Ia (Saha \etal 1994, 1995, 1996; Sandage \etal 1996). From these SNe,
these authors obtained \Ho(B) = 56 $\pm$ 4 \kmsMpc and \Ho(V) = 58 $\pm$ 4 \kmsMpc
(Sandage \etal 1996). Since these values were calculated without including corrections
for the peak luminosity-decline rate relation, a reexamination of the Hubble constant based on
the Cal\'{a}n/Tololo SNe would seem in order. 

Owing to the poor quality of the light curves of SNe 1895B and 1960F
(Schaefer 1995, 1996), we are unable to derive a precise estimate of the decline
rate parameter $\Delta$m$_{15}$(B).  Hence, these two
events are not considered in the remainder of this discussion.
Although adequate photometry is available for SNe 1937C and 1972E, neither was
observed at maximum light, so we must use the same \chisq-minimizing 
template-fitting technique employed for the Cal\'{a}n/Tololo SNe
to estimate their peak magnitudes and decline rates.
SNe 1981B and 1990N, on the other hand, have well-sampled light curves with coverage beginning before
maximum light so that the peak magnitudes and decline rates can be directly measured
from the light curves. To convert these values into absolute magnitudes, we assumed the
{\it true} Cepheid distance moduli measured by Sandage and collaborators and 
the color excesses E(B-V) from Burstein \& Heiles (1984) for
interstellar dust in our own galaxy.
With this approach we wish to make the nearby sample reflect
the properties of the distant sample for which we have not attempted to correct
for possible obscuration in the host galaxies.  Note that the mean
B$_{\rm{MAX}}-$V$_{\rm{MAX}}$ color, corrected for foreground extinction,
of the four calibrating SNe is 0.01$\pm$0.07 (see Table 2),
which is very close to the mean observed color of 0.03$\pm$0.06 of the subsample of
26 Cal\'{a}n/Tololo events.

The relevant data for these four SNe with Cepheid distances are listed in Table 2 
in the following format:\\

\noindent Column (1): SN name.\\
\noindent Column (2): the name of the host galaxy.\\
\noindent Column (3): the host galaxy morphology following the scheme in the Hubble Atlas of
Galaxies (Sandage 1961).\\ 
\noindent Column (4): the true distance modulus of the host galaxy.\\
\noindent Column (5): references to the distance modulus given in column (4).\\
\noindent Column (6), (7) and (8): the apparent B, V, and I peak magnitudes of
the SN as determined from the light curves. Note that the peak magnitudes that
we derived for SN 1937C are brighter by 0.14$^{m}$ in B and 0.18$^{m}$ in V than
those found by Pierce and Jacoby (1995). This difference is ascribed to the
different techniques of fitting the light curve templates to the data.\\
\noindent Column (9): the decline rate parameter $\Delta$m$_{15}$(B).\\
\noindent Column (10): the color excess E(B-V) from Burstein \& Heiles (1984) for
interstellar dust in our own galaxy.\\
\noindent Column (11), (12) and (13): the absolute B, V, and I peak magnitudes
of the SN calculated from the apparent magnitudes, the distance modulus of the
host galaxy, and the assumed color excess.  A standard Galactic reddening law was assumed to
convert the color excess to the extinction in the B, V, and I bands. The error
estimates include uncertainties from the apparent magnitude, distance
modulus, and reddening.\\
\noindent Column (14): references to the SN photometry.\\

\noindent The value of the Hubble constant can be determined from the corrected Hubble
diagrams of the Cal\'{a}n/Tololo SNe along with the absolute magnitudes of the 4 calibrating 
SNe, suitably corrected for their decline rates. From the corrected Hubble relations
[Eqs. 7-9] we have,\\

\vspace{0.2in}
\centerline{log \Ho(B) = 0.2\{M$^{B}$$_{MAX}$ - 0.784($\pm$0.182) [$\Delta$m$_{15}$(B) - 1.1] +28.318($\pm$0.035)\}, (10)}
\vspace{0.2in}
\centerline{log \Ho(V) = 0.2\{M$^{V}$$_{MAX}$ - 0.707($\pm$0.150) [$\Delta$m$_{15}$(B) - 1.1] +28.329($\pm$0.031)\}, (11)}
\vspace{0.2in}
\centerline{log \Ho(I) = 0.2\{M$^{I}$$_{MAX}$ - 0.575($\pm$0.178) [$\Delta$m$_{15}$(B) - 1.1] +28.057($\pm$0.035)\}. (12)}
 
\noindent Table 3 summarizes our calculations for the individual calibrators. In columns (2)-(4)
we give the absolute B, V, and I peak magnitudes of the individual SNe {\it corrected for
the peak luminosity-decline rate relation}, where
the error estimates include uncertainties from the apparent magnitude, foreground reddening, distance
modulus, decline rate, and the slope of the absolute magnitude-decline rate relationship. 
In columns (5)-(7) we give the corresponding values of the Hubble constant obtained from Eqs. (10)-(12).
The errors in \Ho include uncertainties in the
(corrected) absolute magnitudes, the zero point of the Hubble diagram, and the observed scatter
of the individual SNe in the corrected Hubble diagram (0.17$^{m}$ in B, 0.14$^{m}$ in V,
and 0.13$^{m}$ in I). In Table 4 we summarize the error budget of our calculations of the Hubble
constant in the V filter for an individual SN (1972E).
Weighted averages in B, V, and I of the absolute magnitudes and the values of the Hubble constant
for the 4 calibrating SNe are given in the bottom line of Table 3.\\

Averaging the values of the Hubble constant for the three colors gives:\\

\centerline{\Ho = 63.1 $\pm$ 3.4 (internal) $\pm$2.9 (external) \kmsMpc.}

\noindent Since the errors in the individual values of \Ho from B, V, and I are not
all independent, we adopt an error for the mean which is the smallest
of the uncertainties of the individual single-color determinations. The true internal error
is probably smaller than this conservative estimate.
The external error is the uncertainty in the zero point of the Cepheid calibration
(assumed to be $\pm$0.10$^{m}$). The resulting dispersion of $\sigma$=0.5 \kmsMpc 
shows the good agreement among all three determinations of the Hubble constant.\\

Note that these values of \Ho~are in excellent agreement with the preliminary results published
in Paper IV from the 13  Cal\'{a}n/Tololo SNe, namely, \Ho(B)=62$\pm$11 \kmsMpc and
\Ho(V)=63$\pm$8 \kmsMpc.\\

Of the four calibrators, it could be argued that SNe 1981B and 1990N should be given
more weight than SNe 1937C and 1972E for several reasons.  First of all, the maximum 
light magnitudes of 1981B and 1990N were directly observed, whereas we had to estimate
those of 1937C and 1972E through template fitting.  Secondly, and probably more
important, SNe 1981B and 1990N both had decline rates of $\Delta$m$_{15}$(B) $\sim$ 1.1,
which is the reference value we have employed because it is ``typical''
for type~Ia events like those discovered in the
Cal\'{a}n/Tololo survey (see the lower half of Figure 4, or Figure 2 of Paper V).  SNe 1937C and
1972E, on the other hand, were slow-declining events ($\Delta$m$_{15}$(B) $\sim$ 0.9),
and therefore fall at the extreme end of the peak luminosity-decline rate relation 
(see Figure 2 of Paper V).  Use of 1981B and 1990N therefore minimizes the effect of
uncertainties or biases in the slopes of the absolute magnitude-decline rate 
relations.  A weighted average of the absolute magnitudes of Table 3
for SNe 1981B and 1990N alone yields a Hubble constant:\\

\centerline{\Ho = 66.9 $\pm$ 5.2 (internal) $\pm$3.1 (external) \kmsMpc.}

If we ignore the peak luminosity-decline rate relation, the four calibrating SNe~Ia
give \Ho(B) = 55.9 $\pm$ 3.9 \kmsMpc,
\Ho(V) = 56.4 $\pm$ 3.7 \kmsMpc, and
\Ho(I) = 57.1 $\pm$ 5.5 \kmsMpc, which
are in close agreement with the results of Sandage \etal (1996). This similarity is not
surprising since we are using their Cepheid calibrations and the Hubble diagram they
use (Tammann \& Sandage 1995) is heavily weighted by the inclusion of seven distant
Cal\'{a}n/Tololo SNe from Paper IV.  If, however, we use only SNe 1981B and 1990N
as the calibrators for the same reasons given in the previous paragraph, these
values increase to
\Ho(B) = 63.8 $\pm$ 6.5 \kmsMpc,
\Ho(V) = 62.3 $\pm$ 5.9 \kmsMpc, and
\Ho(I) = 60.5 $\pm$ 8.5 \kmsMpc.
Note that the latter values are now very close to the Hubble constant that we derive
when we correct for the absolute magnitude-decline relationship.  {\it This is because
SNe 1981B and 1990N are much more representative (in terms of both decline rate and
luminosity) of the ``average'' SNe~Ia which are included in the Hubble diagrams
of both the present paper and that of Tammann \& Sandage (1995).}  As we have
shown in Paper V, there is now strong evidence that galaxies with younger
stellar population (spirals and irregulars) appear to host the slowest-declining,
and therefore most luminous SNe I.  Hence, the use of 
Pop I objects such as Cepheids to calibrate the zero point of
the SNe Ia Hubble diagram can easily bias the results toward luminous SNe Ia,
{\it unless the absolute magnitude-decline relation is taken into account.}

\subsection{ The Recession Velocities of the Virgo and Fornax Clusters}

      Freedman \etal (1994) have used HST to measure a Cepheid distance to
M100 of 17.1$\pm$2 Mpc ($\mu$=31.16$\pm$0.20)\footnote[3]{ A recent reanalysis 
of these data by the same group gives a distance of 16.1$\pm$1.3 Mpc, or 
$\mu$=31.04$\pm$0.17 (Ferrarese \etal 1996).}. Based on this distance 
and an assumed cosmic recession velocity of 1,404 $\pm$ 80 \kms, this group 
found a value of \Ho=82$\pm$17 \kmsMpc. Whether the Freedman \etal choice of 
1,404 \kms is correct or not rests on the adopted
corrections for the peculiar velocity of Virgo. Estimates of the
extragalactic distance scale based on the Virgo cluster suffer from large
uncertainties due to the possible peculiar velocity of the cluster and cluster
depth. On the other hand, our method of using distant SNe Ia is not 
affected by this problem since the peculiar motions of the 
Cal\'{a}n/Tololo SNe are a small fraction of their redshifts.

We can use our observed Hubble diagram, combined with historical
observations of the SNe Ia in the Virgo  and Fornax clusters to predict their 
recession velocities and relative distances. An earlier attempt to derive
the Virgo distance in this manner was made by Leibundgut \& Tammann (1990)
and Leibundgut \& Pinto (1992); 
more recent photometry and the peak magnitude-decline rate relation
indicate that a rediscussion of this issue is worthwhile.
In Table 5 we summarize the Virgo and Fornax SNe with modern
photometry. The scatter in the photographic light curves for SN 1984A
limits our ability to measure
a precise value of $\Delta$m$_{15}$(B) for this event; we shall assume 
that this SN has the fiducial decline rate value of 1.10 ($\pm$0.10).

Excluding the peculiar event SN 1991bg, we find $<$B$>$=12.16$^m$$\pm$0.20,
$<$V$>$=12.07$^m$$\pm$0.20,\\
and $<$I$>$=12.23$^m$$\pm$0.39 for Virgo and $<$B$>$=12.53$^m$$\pm$0.04, 
$<$V$>$=12.50$^m$$\pm$0.05, and\\
$<$I$>$=12.75$^m$$\pm$0.05 for Fornax.
The results of correcting the peak magnitudes of the individual SNe for 
the magnitude-decline rate relation are shown in Table 6.  Here the quoted
errors for the individual corrected magnitudes include the photometric
and decline rate errors given in Table 5, the errors in the slopes that
we have fitted to the absolute
magnitude-decline rate relation, and the uncertainty in the observed
scatter of the individual SNe in the Hubble diagrams (0.17$^m$ in B,
0.14$^m$ in V, and 0.13$^m$ in I).  Note that the
average $\Delta$m$_{15}$(B) for the Virgo SNe is $\sim$1.10 and therefore the 
mean corrected magnitudes (also shown in Table 6) are very similar to the mean
observed magnitudes given above, while the faster decline rates of the two Fornax SNe
act to make the corrected magnitudes somewhat brighter.
We have not corrected for foreground reddening since the reddening values
are zero for both Virgo and Fornax, based on the maps of Burstein and Heiles
(1982). Regarding possible obscuration in the host galaxies, note that the
mean color of the five Virgo SNe ($<B$-V$>$=0.09$\pm$0.10({\it rms})) and
the two Fornax SNe ($<$B-V$>$=0.03$\pm$0.02({\it rms})) are very close to 
the mean color of the Cal\'{a}n/Tololo SNe 
(($<$B-V$>$=0.03$\pm$0.06({\it rms})). Also, all of the Virgo and
Fornax SNe considered here have B-V $\leq$ 0.20 which is the color cutoff
that we use as an objective criteria to exclude heavily reddened events
in the Cal\'{a}n/Tololo SNe. Hence, we believe that host galaxy reddening in
the Fornax and Virgo SNe is not significantly different than that 
in the distant sample.

Table 6 shows that, in all three colors, the standard deviation of the corrected 
magnitudes for the Virgo SNe is greater than the errors in the individual
magnitudes.  A likely source for this additional dispersion is the
significant depth of the cluster, estimated by Freedman \etal (1994) to
be $\pm$0.35$^m$ and also apparent in the few Cepheid distances of Virgo
cluster galaxies obtained to date (Sandage \etal 1996).  
On the other hand, the standard deviation of the
corrected magnitudes of the two Fornax SNe is much less than the errors
in the individual magnitudes.  Although we expect that depth effects for
Fornax will be small due to the compact nature of the cluster, the extremely
close agreement in the corrected magnitudes of the two SNe must be 
a fortuitous consequence of small number statistics.  Hence, we shall
use the individual errors in the corrected magnitudes to estimate
realistic errors in the mean corrected magnitudes for Fornax, adopting
0.13$^m$ (0.19$^m$/$\sqrt{2}$) in B, 0.11$^m$ (0.16$^m$/$\sqrt{2}$) in V, and
0.10$^m$ (0.15$^m$/$\sqrt{2}$) in I.

If we now apply equations (7), (8) and (9), we derive the cosmic 
recession velocities shown in Table 6.  Restricting
ourselves to the B and V results for Virgo, for which there is data
for all five SNe, the resulting
mean velocity for Virgo is 1,223 \kmsec, to which we conservatively
assign an error 
\footnote[4] {The errors in the
individual recession velocities derived from B and V are at least
partially correlated (e.g., due to the effect of cluster depth), and
so we adopt an error for the mean which is typical of the uncertainties in
the individual single-color determinations.  The true error is probably somewhat
less than this estimate.} of $\pm$115 \kmsec.
The corresponding value for Fornax is
1,342 $\pm$70 \kmsec, where we have used the results in all three colors. 

In Table 7
we show a comparison of these predicted v$_{\rm CMB}$ values with recent
results from the literature. The predicted value for Fornax agrees well
with the measured velocities. The situation for Virgo is more complex, since
there is a wide range of estimates for the correct velocity.
Based on the cosmic recession velocity of Virgo of 1,223 \kmsec estimated 
here we can derive our infall velocity into Virgo. Adopting the observed
heliocentric cluster velocity of 1,050 $\pm$35 \kmsec from 
Bingelli \etal (1993) and a correction of -107 \kmsec to the Local Group
center derived from the Galactic coordinates ($l = 284^\circ$, $b = 74^\circ$)
and the transformation vector given by Lynden-Bell \& Lahav (1988),
we infer an infall of 280 $\pm$ 120 \kmsec.  This value is consistent
with the Aaronson \etal (1982) result of 331 $\pm$ 41 \kmsec and more
recent determinations (e.g., Jerjen \& Tammann 1993, Bureau \etal 1996),
although clearly a larger sample of SNe are needed to improve the
significance of the result.  Note also that there is still disagreement over
the correct value of the heliocentric velocity of Virgo; if we use Huchra's
(1988) preferred value of 1,150 $\pm$51 \kmsec, an infall velocity of 
180 $\pm$126 \kmsec is implied.

From these corrected  mean magnitudes we can also derive relative distance
moduli. We find $\Delta\mu$(B)=0.15$^m$ $\pm$0.24$^m$ and
$\Delta\mu$(V)=0.23$^m$ $\pm$0.23$^m$
for a combined estimate of $\Delta\mu$=0.19$^m$ $\pm$0.23$^m$.
This lies about 1-$\sigma$ from a recent
result from the I-band Tully--Fisher relation of
$\Delta\mu$=--0.06$\pm$0.15 (Bureau \etal 1996).  Again,
the accuracy of the determination will improve considerably when more SNe are
discovered in either cluster.

Finally if we assume the average absolute magnitudes corrected for the peak luminosity-decline
rate relation of the four nearby calibrators given at the bottom of Table 3,
we can make the following predictions:

$\bullet$ The true distance modulus of Virgo is 31.41 $\pm$ 0.22.
Therefore, M100 ($ \mu$=31.04 $\pm$ 0.17)
is most likely on the near side of the cluster.

$\bullet$ The distance modulus of the Fornax cluster is 31.60 $\pm$ 0.15.

Due to the problem of the intrinsic depth of the Virgo cluster
(assumed by Freedman \etal to be $\pm$0.35$^m$ in distance modulus),
Cepheid distances for a number of spirals will be needed before
the first prediction can be verified. 
Hence, the easiest prediction to test is, in fact, the distance to the Fornax cluster since this
cluster is relatively nearby and has a much smaller depth
($<$0.15$^m$). A recent determination of the Cepheid distance to NGC 1365 in
Fornax, yields a distance modulus of 31.30$^m$$\pm$0.23$^m$ (Freedman 1996), in
reasonable agreement with the above determination.

\section{ Conclusions}

\noindent 1) The main goal of the Cal\'{a}n/Tololo survey which we initiated
in 1990 was to address the question of how reliable are Type~Ia SNe as 
extragalactic standard candles.
The Cal\'{a}n/Tololo database has shown that the Hubble diagrams
of SNe Ia are characterized by significant dispersions that range between
$\sim$0.4$^{m}$ in B and $\sim$0.2$^{m}$ in I.  Thus, 
SNe Ia are far from being perfect
standard candles.  Nevertheless, the relatively small dispersion observed
in I is encouraging, and suggests that future work should make more use
of this band.\\

\noindent 2) The Cal\'{a}n/Tololo database has confirmed the finding of
Phillips (1993) from a group of nearby SNe Ia that the luminosities
of these events are correlated with the initial decline rates of their 
light curves (see Paper V),
although this effect seems to be less pronounced than in the original
sample studied by Phillips.
Application of the peak luminosity-decline rate
relationship results in a significant decrease in the scatter
in the BVI Hubble diagrams to levels 0.17$^{m}$-0.13$^{m}$. Thus,
after correcting the apparent magnitudes for the decline rates, SNe Ia
can be used to derive
relative distances with precisions $\sim$7-10\%.\\

\noindent 3) Using published Cepheid distances to SNe 1937C, 1972E, 1981B, and 1990N
to calibrate the zero point of the ``corrected'' BVI Hubble diagrams we
obtain a value for the Hubble constant of \Ho=63.1$\pm$3.4 (internal) $\pm$2.9 (external) \kmsMpc.
The implication of ignoring the peak luminosity-decline rate relation
is to underestimate the Hubble constant by a factor $\sim$10-15\%.\\

\noindent 4) Using the inertial reference frame of the distant SNe Ia,
and the ``corrected'' apparent magnitudes of the five best-observed SNe~Ia 
in the Virgo cluster and two SNe~Ia in the Fornax cluster, we predict
the Hubble flow at Virgo of 1,223$\pm$115 \kms and 1,342$\pm$70 \kms
for Fornax.  The ``corrected'' magnitudes also imply a relative
distance modulus of $\Delta\mu$(Fornax-Virgo) = 0.19$^m$$\pm$0.23$^m$.\\

\noindent 5) Given the high intrinsic brightness, SNe Ia provide a unique
tool for measuring the distances to very distant galaxies and, thus, the
deceleration parameter \qo (e.g., see Perlmutter \etal 1995). This will be,
of course, a challenging task
since a large number of events at redshifts of 0.3-0.5 will have to be
observed. Furthermore, the need to measure the decline rate of such objects
will require large telescopes. With a collective effort, however, such a goal
appears achievable nowadays.\\

                              Acknowledgments

We are extremelly grateful to the large number of CTIO visiting astronomers
and CTIO staff members who gathered data for this project. We also thank
Gary Schmidt and George Jacoby for useful discussions.
This paper has been possible thanks to grant 92/0312 from Fondo Nacional de Ciencias y
Tecnolog\'{i}a (FONDECYT-Chile).
MH acknowledges support provided for this work by the National Science Foundation
through grant number GF-1002-96 from the Association of Universities for Research
in Astronomy, Inc., under NSF Cooperative Agreement No. AST-8947990 and from
Fundaci\'{o}n Andes under project C-12984.
JM acknowledges support by C\'{a}tedra Presidencial 1996/97.

\eject

\centerline{                                References}

\noindent Aaronson, M., Huchra, J., Mould, J., Schechter, P.L., \& Tully, R.B. 1982,
     ApJ, 258, 64

\noindent Barbon, R., Ciatti, F., \& Rosino, L. 1982, A\&A, 116, 35

\noindent Barbon, R., Iijima, T., \& Rosino, L. 1989, A\&A, 220, 83

\noindent Bingelli, B., Popescu, C.C., \& Tammann, G.A. 1993, A\&AS, 98, 275

\noindent Bureau, M., Mould, J.R., \& Stavely-Smith, L. 1996, ApJ, 463, 60

\noindent Burstein, D., \& Heiles, C. 1982, AJ, 87, 1165

\noindent Burstein, D., \& Heiles, C. 1984, ApJS, 54, 33

\noindent Buta, R.J., \& Turner, A. 1983, PASP, 95, 72

\noindent Ferrarese, L., \etal 1996, ApJ, 464, 568

\noindent Filippenko, A.V., \etal 1992, AJ, 104, 1543

\noindent Freedman, W.L., \etal 1994, Nature, 371, 757

\noindent Freedman, W.L. 1996, in proceedings of {\it The Extragalactic Distance Scale}, 
     STScI Spring meeting, ed. M. Livio, in preparation

\noindent Hamuy, M., Phillips, M.M., Maza, J., Wischnjewsky, M., Uomoto, A., 
     Landolt, A.U., \& Khatwani, R. 1991, AJ, 102, 208

\noindent Hamuy, M., \etal 1993a, AJ, 106, 2392 (Paper I)

\noindent Hamuy, M., Phillips, M.M., Wells, L.A., \& Maza, J. 1993b, PASP, 105, 787

\noindent Hamuy, M., \etal 1994, AJ, 108, 2226 (Paper III)

\noindent Hamuy, M., Phillips, M.M., Maza, J., Suntzeff, N.B., Schommer, R.A.,
     \& Avil\'{e}s, R. 1995, AJ, 109, 1 (Paper IV)

\noindent Hamuy, M., Phillips, M.M., Schommer, R.A., Suntzeff, N.B., Maza, J.,
     \& Avil\'{e}s, R. 1996a, AJ, this volume (Paper V)

\noindent Hamuy, M., \etal 1996b, AJ, this volume (Paper VII)
   
\noindent Hamuy, M., \etal 1996c, AJ, this volume (Paper VIII)

\noindent Held, E.V., \& Mould, J.R. 1994, AJ 107, 1307

\noindent Huchra, J.P. 1988, in {\it The Extragalactic Distance Scale},
     ASP Conference Series, Vol. 4, ed. S. van den Bergh \&
     C.J. Pritchet (Provo: Brigham Young Univ. Print Services), 257

\noindent Jerjen, H., \& Tammann, G.A. 1993, A\&A, 276, 1

\noindent Kimeridze, G.N., \& Tsvetkov, D.Y. 1986, Ap, 25, 513

\noindent Leibundgut, B., \& Tammann, G.A. 1990, A\&A, 230, 81

\noindent Leibundgut, B., \& Pinto, P.A. 1992, ApJ, 401, 49

\noindent Leibundgut, B., \etal 1993, AJ, 105, 301

\noindent Lira, P. 1995, MS Thesis, Universidad de Chile.

\noindent Lynden-Bell, D., \& Lahav, O. 1988, in {\it Large-Scale Motions in the Universe}
     ed. V.C. Rubin \& G.V. Coyne (Princeton: Princeton Univ. Press), 199

\noindent Marzke, R. O., Geller, M. J., da Costa L. N., \& Huchra, J. P., 1995, AJ, 110, 477

\noindent Maza, J., Hamuy, M., Phillips, M.M., Suntzeff, N.B., \& Avil\'{e}s, R. 1994,
     ApJ, 424, L107 (Paper II)

\noindent Nugent, P., Phillips, M., Baron, E., Branch, D., \& Hauschildt, P. 1995,
     ApJ, 455, L147

\noindent Perlmutter, S., \etal 1995, ApJ, 440, L41

\noindent Phillips, M.M., \etal 1987, PASP, 99, 592

\noindent Phillips, M.M. 1993, ApJ, 413, L105
 
\noindent Phillips, M.M., \& Eggen, O.J. 1996, in preparation

\noindent Pierce, M.J., \& Jacoby, G.H. 1995, AJ, 110, 2885

\noindent Saha, A., Labhardt, L., Schwengeler, H., Macchetto, F.D., Panagia, N.,
     Sandage, A., \& Tammann, G.A. 1994, ApJ, 425, 14

\noindent Saha, A., Sandage, A., Labhardt, L., Schwengeler, H., Tammann, G.A., 
     Panagia, N., \& Macchetto, F.D. 1995, ApJ, 438, 8

\noindent Saha, A., Sandage, A., Labhardt, L., Tammann, G.A., Macchetto, F.D.,
     \& Panagia, N. 1996, ApJ, in press

\noindent Sandage, A. 1961, {\it The Hubble Atlas of Galaxies} (Carnegie Institution of 
     Washington, Washington DC)

\noindent Sandage, A., Saha, A., Tammann, G.A., Labhardt, L., Panagia, N., \&
     Macchetto, F.D. 1996, ApJ, 460, L15

\noindent Schaefer, B.E. 1995, ApJ, 447, L13

\noindent Schaefer, B.E. 1996, ApJ, 460, L19

\noindent Smith, R.C., \etal 1996, in preparation

\noindent Smoot, G.F., \etal 1992, ApJ, 396, L1

\noindent Suntzeff, N.B. \etal 1996a, in preparation

\noindent Suntzeff, N.B. \etal 1996b, in preparation

\noindent Tammann, G.A., \& Sandage, A. 1995, ApJ, 452, 16

\noindent Tsvetkov, D.Y. 1982, SvAL, 8, 115

\noindent Vaughan, T.E., Branch, D., Miller, D.L., \& Perlmutter, S. 1995, ApJ, 439, 558

\eject

\centerline{                             Figure Captions}

\noindent Figure 1. The Hubble diagrams in B, V, and I for the 29 Cal\'{a}n/Tololo SNe Ia.\\

\noindent Figure 2. The absolute B, V, and I magnitudes of the Cal\'{a}n/Tololo sample of
29 SNe Ia, plotted as a function of their B$_{\rm{MAX}}-$V$_{\rm{MAX}}$ colors. The
Galactic reddening vectors (with slopes of 4.1 in B, 3.1 in V, and 1.85 in I)
corresponding to E(B-V)=0.2 are indicated.\\

\noindent Figure 3. The absolute B, V, and I magnitudes of the Cal\'{a}n/Tololo SNe Ia
plotted as a function of redshift. The vertical line at log cz = 4 (10,000 \kms)
illustrates the separation of the sample into two groups.\\

\noindent Figure 4. (top) The morphological types of their host galaxies of the 
Cal\'{a}n/Tololo SNe Ia plotted as a function of redshift. Note that 
the ratio of spirals/nonspirals is significantly higher in the 
nearby sample (log cz $<$ 4) than in the more
distant group (log cz $>$ 4). (bottom) The decline rate ( $\Delta$m$_{15}$(B))
plotted vs redshift. Note the wide spread in decline rates in the nearby
sample, indicative of the full range of morphological types seen in
the top panel.\\

\noindent Figure 5. (top panel) The Hubble diagram in V for the SNe Ia in the Cal\'{a}n/Tololo
sample with B$_{\rm{MAX}}-$V$_{\rm{MAX}}$ $\leq$ 0.20. (bottom panel) The Hubble diagram
for the same 26 events after correction for the peak luminosity-decline rate dependence.\\

\eject
\pagestyle{empty}
\evensidemargin=-0.5in
\oddsidemargin=-0.5in

\tiny
\begin{tabular}{lccccccccll}
\multicolumn{11}{c}{\bf Table 1. Colors and Magnitudes of the Cal\'{a}n/Tololo Supernovae Ia} \\
&&&&&&&&&&\\
(1) & (2) & (3) & (4) & (5) & (6) & (7) & (8) & (9) & (10) & (11) \\
&&&&&&&&&&\\
\hline\hline\\
                                                                                   
 SN & log(cz) & B$_{MAX}$ & V$_{MAX}$ & I$_{MAX}$ & B$_{\rm{MAX}}-$V$_{\rm{MAX}}$ & M$^B_{\rm{MAX}}$ & M$^V_{\rm{MAX}}$ & M$^I_{\rm{MAX}}$ & $\Delta$m$_{15}$(B) & EW(NaID) \\
    & CMB & & & & &  +5log(H$_0$/65)& +5log(H$_0$/65)& +5log(H$_0$/65) & &  [\AA] \\
& & & & & & & & & & \\
\hline\\

90O  & 3.958(28) & 16.32(10) & 16.31(08) & 16.70(09) & 0.01(05) & -19.40(17)  & -19.41(16)  & -19.02(17)  & 0.96(10)  &  $<$0.2 \\ 
90T  & 4.080(22) & 17.16(21) & 17.12(16) & 17.35(15) & 0.04(10) & -19.17(24)  & -19.21(19) &  -18.98(19)  & 1.15(10)  &  1.0  \\
90Y  & 4.066(22) & 17.70(21) & 17.37(16) & 17.61(15) & 0.33(10) & -18.56(24)  & -18.89(20)  & -18.65(19)  & 1.13(10)  &  $<$4.0  \\
90af & 4.178(17) & 17.87(07) & 17.82(06) &  ---      &  0.05(03) &  -18.95(11) & -19.00(11) &     ---      & 1.56(05)  & $<$0.2  \\
91S  & 4.223(16) & 17.81(21) & 17.77(16) & 18.07(15) & 0.04(10)  & -19.24(22)  & -19.28(18)  & -18.98(17)  & 1.04(10)  & $<$0.4  \\
91U  & 3.992(28) & 16.40(21) & 16.34(16) & 16.52(15) & 0.06(10)  & -19.49(25)  & -19.55(21)  & -19.37(20)  & 1.06(10)  & 0.8  \\
91ag & 3.616(63) & 14.62(14) & 14.54(15) & 14.86(19) &  0.08(05) &  -19.40(35) & -19.48(35) &  -19.16(37) &  0.87(10) &  0.0  \\
92J  & 4.137(20) & 17.70(21) & 17.58(16) & 17.84(15) & 0.12(10)  & -18.92(23)  & -19.04(19)  & -18.78(18)  & 1.56(10)  & $<$0.2  \\
92K  & 3.523(92) & 15.83(21) & 15.09(16) & 14.94(15) & 0.74(10)  & -17.72(44)  & -18.46(42)  & -18.61(42)  & 1.93(10)  & $<$0.2  \\
92P  & 3.897(35) & 16.08(07) & 16.11(06) & 16.39(06) & -0.03(03) &   -19.34(18) &-19.31(18) &  -19.03(18) &   0.87(10) & 1.2  \\
92ae & 4.351(12) & 18.62(12) & 18.51(08) &  ---      &  0.11(05) &   -19.07(13) &-19.18(10) &    ---     &  1.28(10) &  $<$0.5  \\
92ag & 3.891(36) & 16.41(08) & 16.28(07) & 16.41(06) &  0.13(05) &   -18.98(19) &-19.11(18) &  -18.98(18) &  1.19(10) & 1.1  \\
92al & 3.627(60) & 14.60(07) & 14.65(06) & 14.94(06) &  -0.05(03) &  -19.47(32) &-19.42(31) &  -19.13(31) &  1.11(05) & $<$0.05 \\
92aq & 4.481(09) & 19.45(09) & 19.35(07) & 19.77(09) &  0.10(05)  & -18.89(10)  &-18.99(08)  & -18.57(10)  & 1.46(10)  & $<$1.1  \\
92au & 4.260(14) & 18.21(21) & 18.16(16) & 18.41(15) &  0.05(10)  & -19.03(22)  &-19.08(18)  & -18.83(17)  & 1.49(10)  & $<$1.3 \\
92bc & 3.774(44) & 15.16(07) & 15.24(06) & 15.58(05) &  -0.08(03) &  -19.64(23) &-19.56(23) &  -19.22(22) &  0.87(05) &  $<$0.2  \\
92bg & 4.030(25) & 16.72(08) & 16.76(07) & 17.04(06) &  -0.04(05) &  -19.36(15) &-19.32(14) &  -19.04(14) &  1.15(10) & $<$0.3  \\
92bh & 4.131(20) & 17.70(08) & 17.62(06) & 17.80(06) &  0.08(05)  & -18.89(13)  &-18.97(11)  & -18.79(11)  & 1.05(10)  & $<$0.5  \\
92bk & 4.240(15) & 18.11(10) & 18.11(07) & 18.31(06) &  0.00(05)  & -19.03(12)  &-19.03(10)  & -18.83(10)  & 1.57(10)  & $<$0.4 \\
92bl & 4.110(20) & 17.36(08) & 17.36(07) & 17.64(06) &  0.00(05)  & -19.13(13)  &-19.13(12)  & -18.85(12)  & 1.51(10)  & $<$0.5  \\
92bo & 3.736(46) & 15.86(07) & 15.85(06) & 15.97(05) &  0.01(03)  & -18.76(25)  &-18.77(25)  & -18.65(24)  & 1.69(05)  & $<$0.2  \\
92bp & 4.374(11) & 18.41(07) & 18.46(06) & 18.78(06) &  -0.05(05) &  -19.40(09) &-19.35(08) &  -19.03(08) &  1.32(10) &  $<$1.1  \\
92br & 4.420(10) & 19.38(17) & 19.34(10) &  ---      &  0.04(05)  & -18.66(18)  &-18.70(11)  &   ---       & 1.69(10)  & $<$4.0 \\
92bs & 4.279(14) & 18.37(09) & 18.30(07) &  ---      &  0.07(05)  & -18.96(11)  &-19.03(10)  &   ---       & 1.13(10)  & $<$3.1  \\
93B  & 4.326(13) & 18.53(11) & 18.41(09) & 18.70(10) & 0.12(05)   & -19.04(13)  &-19.16(11)  & -18.87(12)  & 1.04(10)  & $<$0.5 \\
93H  & 3.872(37) & 16.84(08) & 16.61(06) & 16.55(06) & 0.23(05)   & -18.45(19)  &-18.68(19)  & -18.74(19)  & 1.69(10)  &  1.2 \\
93O  & 4.193(17) & 17.67(07) & 17.76(06) & 17.99(06) & -0.09(03)  & -19.23(11)  &-19.14(10)  & -18.91(10)  & 1.22(05)  &  $<$0.1 \\
93ag & 4.177(18) & 17.72(08) & 17.69(06) & 18.01(06) &  0.03(05)  & -19.10(12)  &-19.13(11)  & -18.81(11)  & 1.32(10)  &  ? \\
93ah & 3.935(29) & 16.33(21) & 16.37(16) & 16.68(15) &  -0.04(10) &  -19.28(26) &-19.24(22) &  -18.93(21) & 1.30(10)  &   ? \\
\hline\hline\\
             
\end{tabular}
\eject

\tiny
\evensidemargin=-0.7in
\oddsidemargin=-0.7in
\begin{tabular}{ccccccccccccll}
\multicolumn{14}{c}{\bf Table 2. Nearby SNe Sample     }\\
&&&&&&&&&&&&&\\
(1) & (2) & (3) & (4) & (5) & (6) & (7) & (8) & (9) & (10) & (11) & (12) & (13) & (14) \\
&&&&&&&&&&&&&\\
\hline\hline\\
&&&&Distance&&&&&&&&\\
 SN & Host & Morph. & Distance  & Modulus & B$_{MAX}$ & V$_{MAX}$ & I$_{MAX}$ & $\Delta$m$_{15}$(B)  & E(B-V)& M$^B_{\rm{MAX}}$& M$^V_{\rm{MAX}}$ &  M$^I_{\rm{MAX}}$ & Phot \\
&Galaxy& Type &Modulus& Ref$^{a}$&&&&&&&&&Ref$^{b}$\\
&&&&&&&&&&&&&\\
\hline
&&&&&&&&&&&&&\\
1937C & IC 4182  & S/Irr  & 28.36(09) & 1 &  8.80(09) &  8.82(11) &   ---     &  0.87(10) & 0.00(02) & -19.56(15) & -19.54(16) &            & 1 \\
1972E & NGC 5253 & Irr    & 27.97(07) & 2 &  8.49(14) &  8.49(15) &  8.80(19) &  0.87(10) & 0.05(02) & -19.69(18) & -19.64(18) & -19.26(21) & 2 \\
1981B & NGC 4536 & Sc     & 31.10(13) & 3 & 12.03(03) & 11.93(03) &   ---     &  1.10(05) & 0.00(02) & -19.07(16) & -19.17(15) &            & 3,4,5 \\
1990N & NGC 4639 & Sb     & 32.00(23) & 4 & 12.74(03) & 12.72(03) & 12.95(05) &  1.07(05) & 0.00(02) & -19.26(25) & -19.28(24) & -19.05(24) & 6 \\ 
&&&&&&&&&&&&&\\
\hline\hline\\
\end{tabular}

 $^{a}$References.-
 $^{1}$Saha \etal (1994);
 $^{2}$Saha \etal (1995);
 $^{3}$Saha \etal (1996);
 $^{4}$Sandage \etal (1996).

 $^{b}$References.-
 $^{1}$Pierce \& Jacoby (1995);
 $^{2}$Phillips \& Eggen (1996);
 $^{3}$Buta \& Turner (1983);
 $^{4}$Barbon \etal (1982);
 $^{5}$Tsvetkov (1982);
 $^{6}$Lira (1995).

\eject
\small
\evensidemargin=-0.5in
\oddsidemargin=-0.5in
\begin{tabular}{cccccll}
\multicolumn{7}{c}{\bf Table 3. Corrected Absolute Magnitudes and Values of \Ho}\\
&&&&&&\\
(1) & (2) & (3) & (4) & (5) & (6) & (7) \\
&&&&&&\\
\hline\hline\\
&&&&&&\\
 SN & M$^B_{\rm{MAX,corr}}$& M$^V_{\rm{MAX,corr}}$ &  M$^I_{\rm{MAX,corr}}$ & \Ho(B) & \Ho(V) & \Ho(I) \\
&&&&&&\\
\hline
&&&&&&\\
1937C       &-19.38(18)     &-19.38(17)     &   ---         &  61.3(7.0)&  61.7(6.4) &          \\
1972E       &-19.50(20)     &-19.47(19)     &-19.13(22)     &  57.9(7.0)&  59.1(6.6) & 61.0(7.2)\\
1981B       &-19.07(16)     &-19.17(15)     &   ---         &  70.7(7.7)&  67.9(6.5) &          \\
1990N       &-19.24(25)     &-19.26(24)     &-19.03(24)     &  65.5(9.2)&  65.2(8.5) & 63.8(8.1)\\ 
&&&&&&\\
Weighted &&&&&&\\
Average   &-19.28(10)     &-19.31(09)     &-19.08(16)     &  63.3(3.8)&  63.3(3.4) & 62.2(5.4)\\ 
          & $\sigma$=0.20 & $\sigma$=0.14 & $\sigma$=0.07 &  $\sigma$=5.6 & $\sigma$=4.0 & $\sigma$=2.0\\
&&&&&&\\
\hline\hline\\
\end{tabular}

Notes:\\
\indent M$^B_{\rm{MAX,corr}}$ = M$^B_{\rm{MAX}}$ - 0.784 [$\Delta$m$_{15}$ - 1.1]\\
\indent M$^V_{\rm{MAX,corr}}$ = M$^V_{\rm{MAX}}$ - 0.707 [$\Delta$m$_{15}$ - 1.1]\\
\indent M$^I_{\rm{MAX,corr}}$ = M$^I_{\rm{MAX}}$ - 0.575 [$\Delta$m$_{15}$ - 1.1]\\

\eject
\small
\evensidemargin=-0.0in
\oddsidemargin=-0.0in
\begin{tabular}{ccc}
\multicolumn{3}{c}{\bf Table 4. Error Budget in the Calculation of \Ho(V)}\\
&&\\
&&\\
\hline\hline\\
 &                                   & SN 1972E   \\
 & Error Source                      & Error(mag) \\
&&\\
\hline
&&\\
1 & Apparent V Magnitude                                & 0.15$^{m}$  \\
2 & Reddening (=3.1*$\sigma$$_{E(B-V)}$)                 & 0.06$^{m}$  \\
3 & Distance Modulus                                    & 0.07$^{m}$  \\
\hline
&&\\
4  & Absolute V Magnitude (=(1)+(2)+(3))                & 0.18$^{m}$    \\
5  & M$_{MAX}$\ vs $\Delta$m$_{15}$ slope (=0.15*[$\Delta$m$_{15}$-1.1]) & 0.03$^{m}$  \\
6  & $\Delta$m$_{15}$ (=0.10$^{m}$*0.707)                & 0.07$^{m}$  \\
\hline
&&\\
7  & Corrected Absolute V Magnitude (=(4)+(5)+(6))      & 0.19$^{m}$  \\
8  & Zero Point of Hubble Diagram                       & 0.031$^{m}$ \\
9  & Scatter of Hubble Diagram                          & 0.14$^{m}$  \\
\hline
&&\\
   & Total (=(7)+(8)+(9))                               & 0.24$^{m}$ \\ 
   &                                                    & =6.6 \kmsMpc \\
&&\\
\hline\hline\\
\end{tabular}

\eject

\small
\evensidemargin=-0.5in
\oddsidemargin=-0.5in
\begin{tabular}{cccccccl}
\multicolumn{7}{c}{\bf Table 5. SNe in Virgo and Fornax with Modern Photometry}\\
&&&&&&\\
(1) & (2) & (3) & (4) & (5) & (6) & (7)   \\
&&&&&&\\
\hline\hline\\
&&&&&&\\
 SN & Host & B$_{obs}$ &V$_{obs}$   &I$_{obs}$   & $\Delta$m$_{15}(B)$   & Photometry   \\
&Galaxy&  &&&&References\\
&&&&&&\\
\hline
&&&&&&\\
&&&Virgo&&&&\\
&&&&&&\\
1981B & NGC 4536 & 12.03(03) & 11.93(03) &  ---       & 1.10(05)  & 1,2,3 \\
1984A & NGC 4419 & 12.50(10) & 12.30(10) &  ---       &  ---      &  4,5 \\
1990N & NGC 4639 & 12.74(03) & 12.72(03) & 12.95(05)  & 1.07(05)  &  6 \\ 
1991T & NGC 4527 & 11.69(03) & 11.51(03) & 11.62(03)  & 0.94(05)  &  6 \\ 
1991bg& NGC 4374 & 14.76(10) & 13.97(05) & 13.51(05)  & 1.93(10)  &  7,8 \\  
1994D & NGC 4526 & 11.86(03) & 11.90(03) & 12.11(03)  & 1.32(05)  &  9 \\
&&&&&&\\
&&&Fornax&&&\\
&&&&&&\\
80N & NGC 1316 & 12.49(03) & 12.44(03) & 12.70(04)  &  1.28(04)  & 10 \\
92A & NGC 1380 & 12.57(03) & 12.55(03) & 12.80(03)  &  1.47(05)  & 11 \\
\hline\hline\\
\end{tabular}

 References.-
 $^{1}$Buta \& Turner (1983);
 $^{2}$Barbon \etal (1982);
 $^{3}$Tsvetkov (1982);\\
 $^{4}$Barbon \etal (1989);
 $^{5}$Kimeridze \& Tsvetkov (1986);
 $^{6}$Lira (1995);
 $^{7}$Filippenko \etal (1992);\\
 $^{8}$Leibundgut \etal (1993);
 $^{9}$Smith \etal (1996);
 $^{10}$Hamuy \etal (1991);
 $^{11}$Suntzeff \etal (1996b).
\eject

\small
\evensidemargin=-0.5in
\oddsidemargin=-0.5in
\begin{tabular}{cccc}
\multicolumn{4}{c}{\bf Table 6. Corrected Magnitudes for SNe in Virgo and Fornax}\\
&&&\\
(1) & (2) & (3) & (4) \\
&&&\\
\hline\hline\\
&&&\\
 SN & B$_{corr}$ & V$_{corr}$ & I$_{corr}$ \\
&&&\\
\hline
&&&\\
Virgo:&&&\\
&&&\\
1981B &  12.03(18) & 11.93(15) &  ---       \\
1984A &  12.50(21) & 12.30(19) &  ---       \\
1990N &  12.76(18) & 12.74(15) & 12.97(14)  \\ 
1991T &  11.82(18) & 11.62(15) & 11.71(14)  \\ 
1994D &  11.69(18) & 11.74(15) & 11.98(14)  \\
&&&\\
Average & 12.16(20) & 12.07(20) & 12.22(38) \\
& $\sigma$=0.46 & $\sigma$=0.46 & $\sigma$=0.66 \\
&&&\\
Velocity(CMB) & 1,246(119) \kmsec & 1,200(114) \kmsec & 1,137(200) \kmsec \\
&&&\\
&&&\\
Fornax:&&&\\
&&&\\
80N &  12.35(18) & 12.31(15) & 12.60(14)  \\
92A &  12.28(19) & 12.29(16) & 12.59(15)  \\
&&&\\
Average & 12.31(03) & 12.30(01) & 12.59(01) \\
& $\sigma$=0.05 & $\sigma$=0.02 & $\sigma$=0.01 \\
&&&\\
Velocity(CMB) & 1,339(83) \kmsec & 1,337(69) \kmsec & 1,348(68) \kmsec\\
&&&\\
\hline\hline\\
\end{tabular}

Notes:\\
\indent B$_{corr}$ = B$_{obs}$ - 0.784 [$\Delta$m$_{15}$ - 1.1]\\
\indent V$_{corr}$ = V$_{obs}$ - 0.707 [$\Delta$m$_{15}$ - 1.1]\\
\indent I$_{corr}$ = I$_{obs}$ - 0.575 [$\Delta$m$_{15}$ - 1.1]\\

\eject

\rm
\small
\begin{tabular}{clcc}
\multicolumn{4}{c}{\bf Table 7. Recessional Velocities of the Virgo and Fornax
 Clusters}\\
&&&\\
\hline\hline\\
Cluster & Predicted & Previous & Reference\\
        & v$_{CMB}$ & Values & \\
 & [\kmsec] & [\kmsec] &\\
& &  & \\
\hline\\
Virgo & 1,223 $\pm$115  & 1,179 & Jerjen \& Tammann 1993 \\
     &                & 1,404 & Huchra 1988\\
&&&\\
Fornax & 1,342 $\pm$70  &  1,338 & Jerjen \& Tammann 1993\\
       &               &  1,354 & Held \& Mould 1994  \\
\hline\hline\\
\end{tabular}

\eject
\begin{figure}
\psfull
\psfig{figure=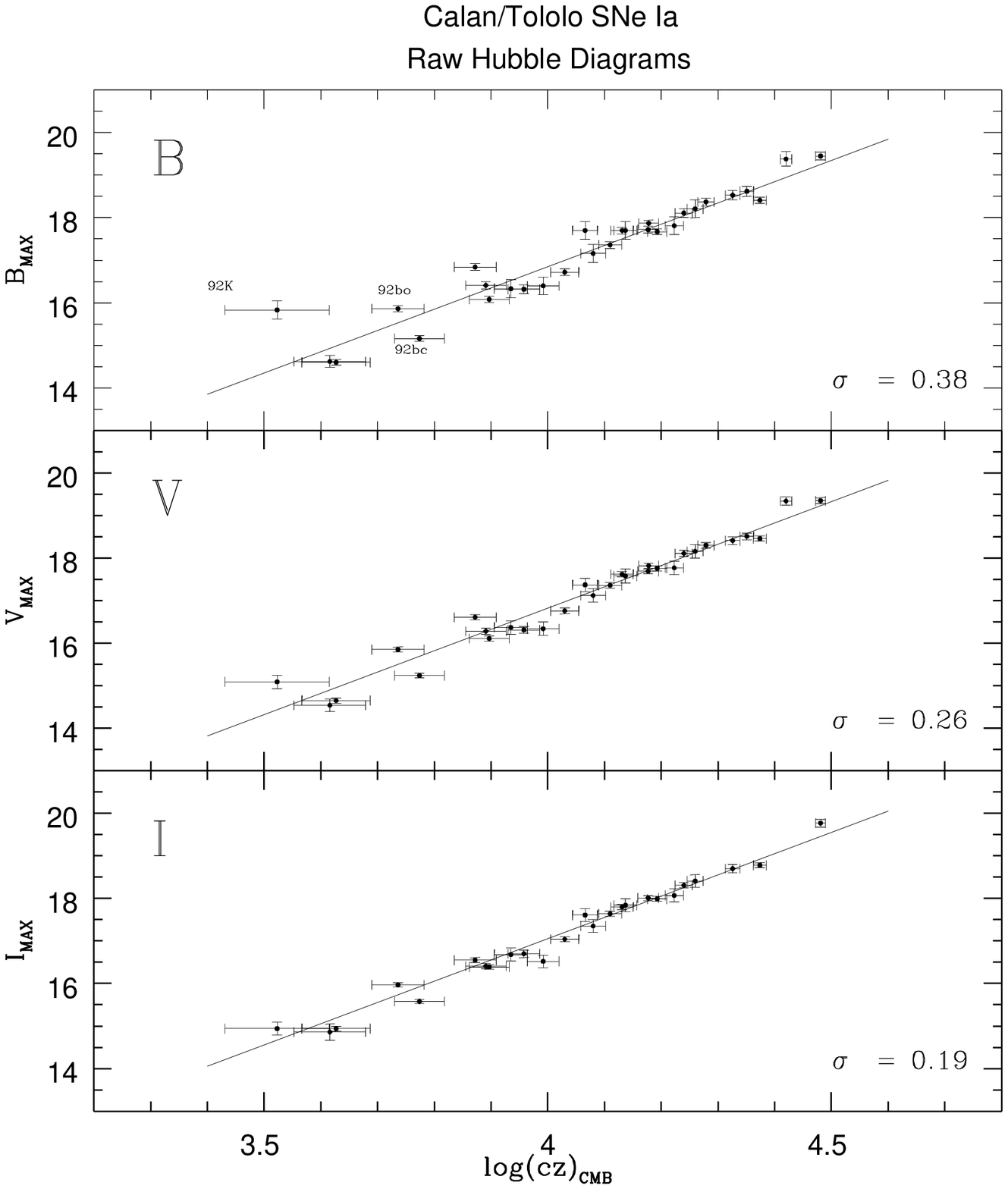}
\caption{The Hubble diagrams in B, V, and I for the 29 Cal\'{a}n/Tololo SNe Ia.}
\end{figure}

\eject
\begin{figure}
\psfull
\psfig{figure=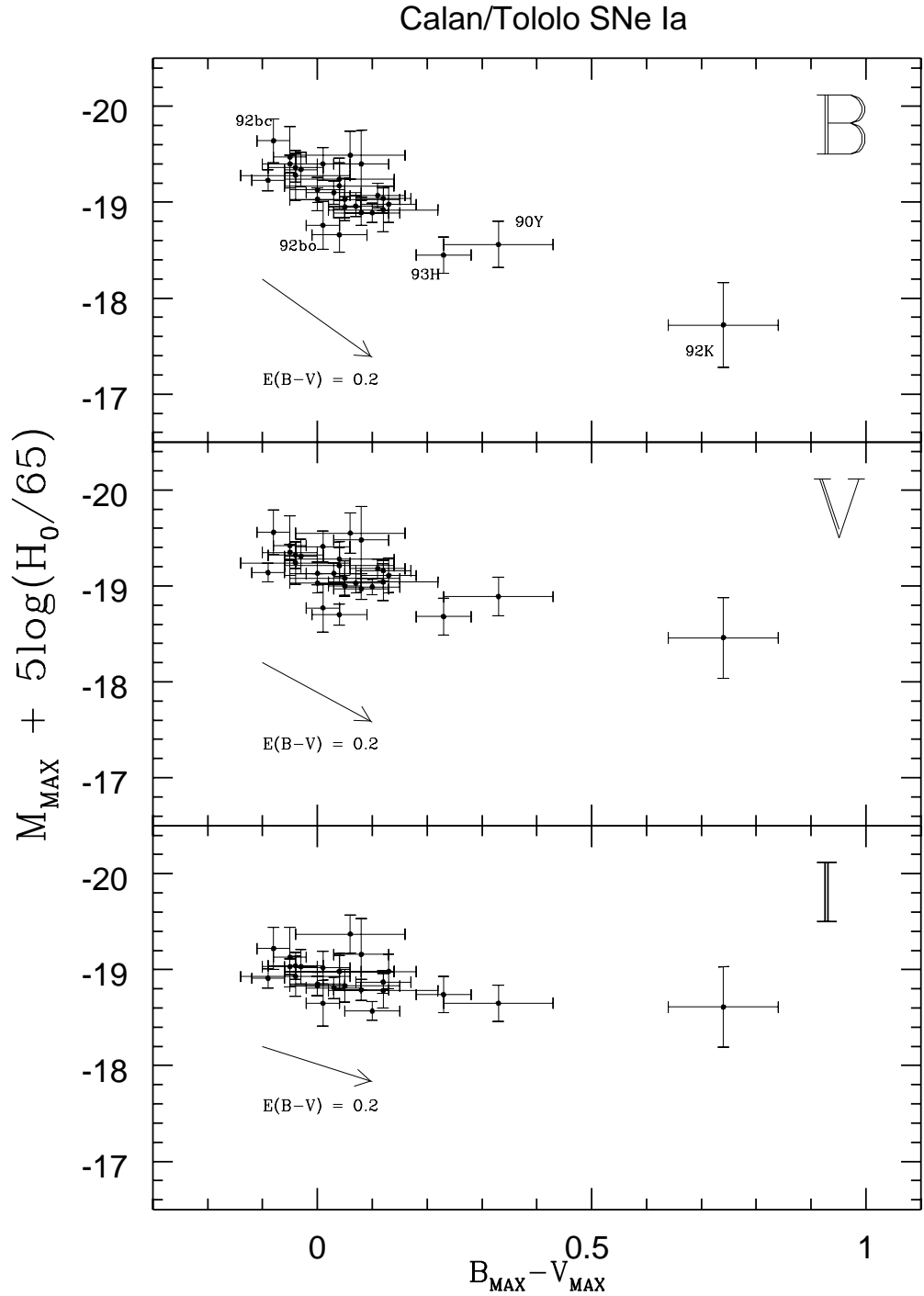}
\caption{The absolute B, V, and I magnitudes of the Cal\'{a}n/Tololo sample of
29 SNe Ia, plotted as a function of their B$_{\rm{MAX}}-$V$_{\rm{MAX}}$ colors. The
Galactic reddening vectors (with slopes of 4.1 in B, 3.1 in V, and 1.85 in I)
corresponding to E(B-V)=0.2 are indicated.}
\end{figure}

\eject
\begin{figure}
\psfull
\psfig{figure=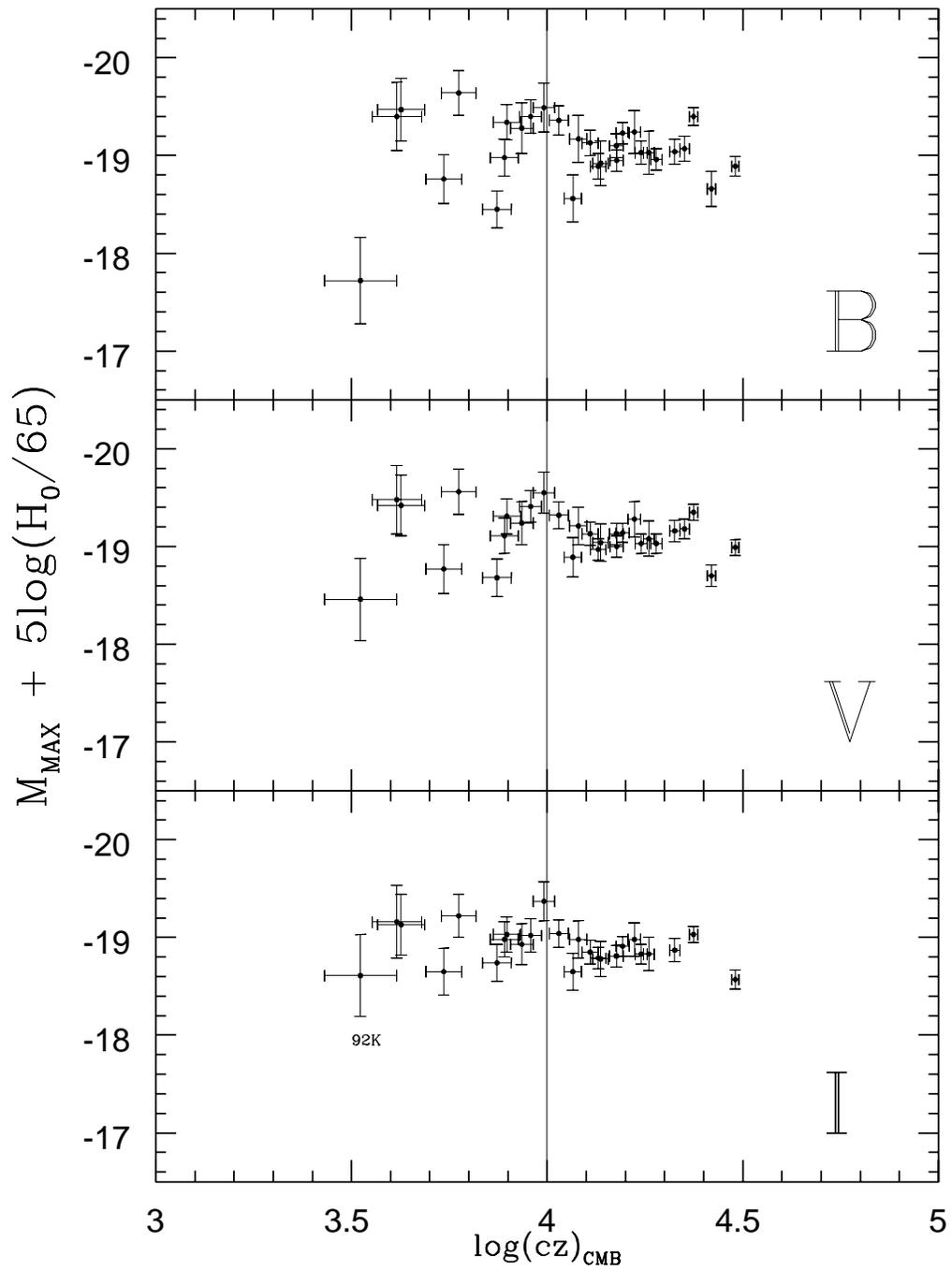}
\caption{The absolute B, V, and I magnitudes of the Cal\'{a}n/Tololo SNe Ia
plotted as a function of redshift. The vertical line at log cz = 4 (10,000 \kms)
illustrates the separation of the sample into two groups.}
\end{figure}

\eject
\begin{figure}
\psfull
\psfig{figure=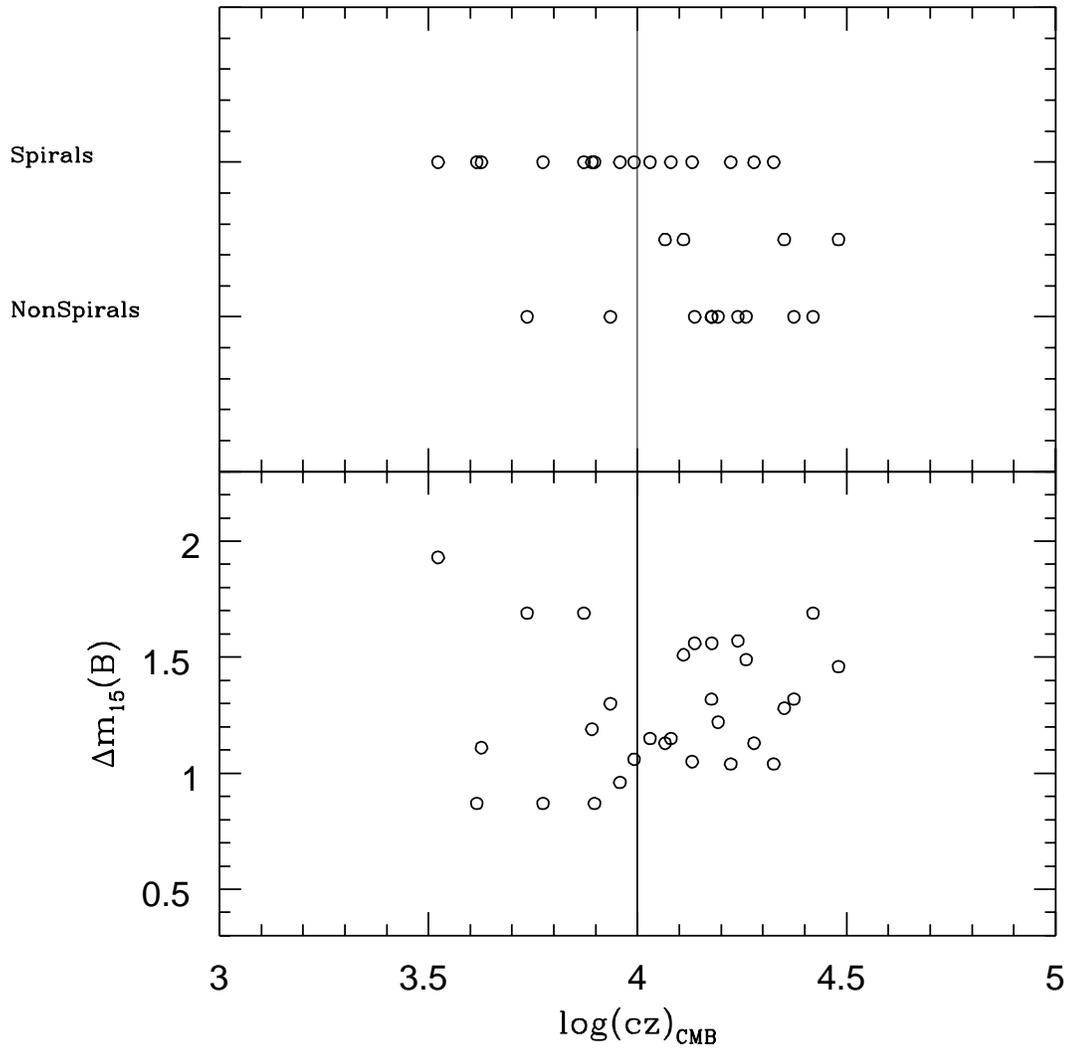}
\caption{(top) The morphological types of their host galaxies of the
Cal\'{a}n/Tololo SNe Ia plotted as a function of redshift. Note that
the ratio of spirals/nonspirals is significantly higher in the
nearby sample (log cz $<$ 4) than in the more
distant group (log cz $>$ 4). (bottom) The decline rate ( $\Delta$m$_{15}$(B))
plotted vs redshift. Note the wide spread in decline rates in the nearby
sample, indicative of the full range of morphological types seen in
the top panel.}
\end{figure}

\eject
\begin{figure}
\psfull
\psfig{figure=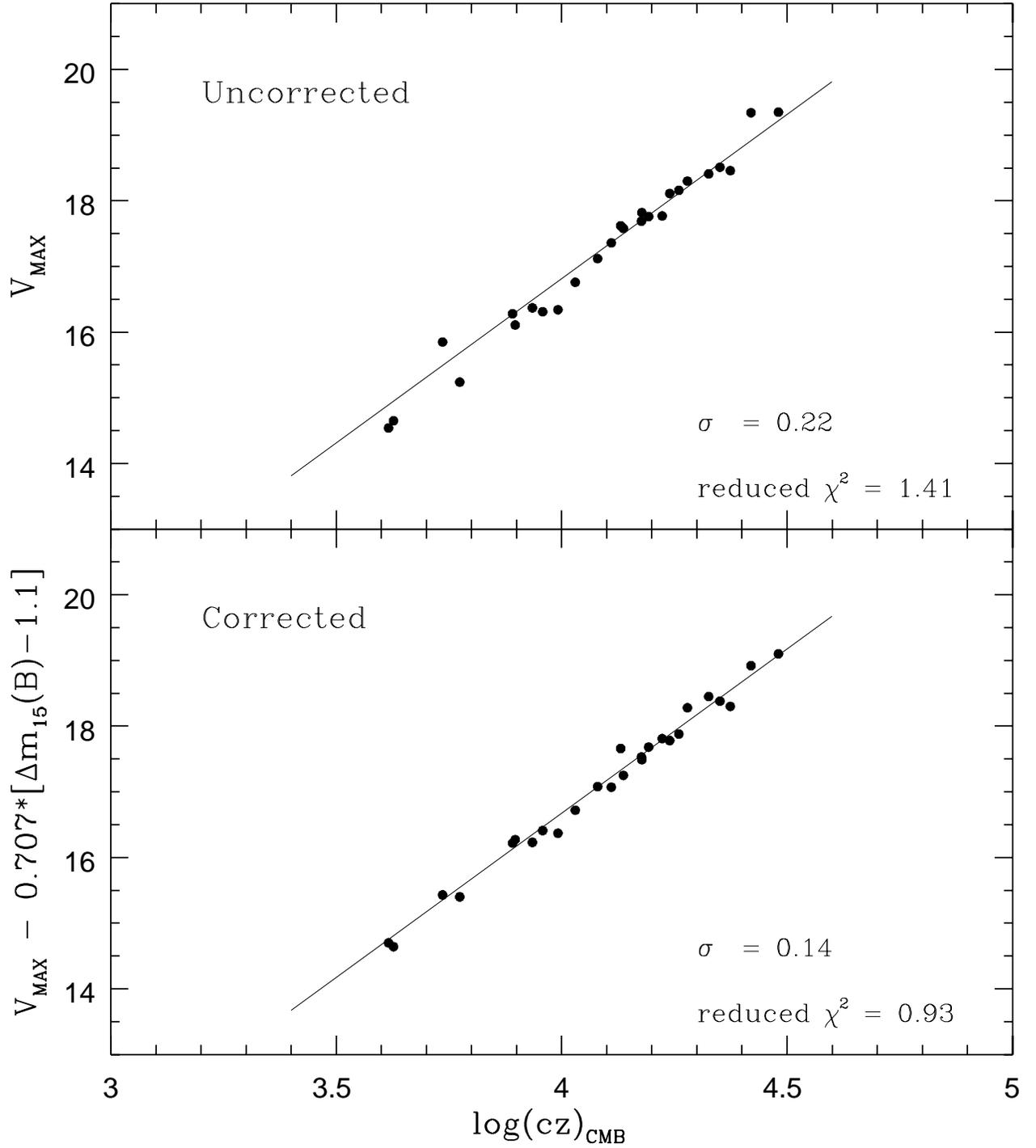}
\caption{(top panel) The Hubble diagram in V for the SNe Ia in the Cal\'{a}n/Tololo
sample with B$_{\rm{MAX}}-$V$_{\rm{MAX}}$ $\leq$ 0.20. (bottom panel) The Hubble diagram
for the same 26 events after correction for the peak luminosity-decline rate dependence.}
\end{figure}

\end{document}